# Extrasolar planets: from dust to new worlds

Barbara Ercolano

*Ludwig-Maximilians-Universitaet, University Observatory Munich (USM)
Scheinerstr 1, D-81679, Munich, Germany*

**Thousands of exoplanets have been discovered and the search for life outside Earth is at the forefront of astrophysical research. The planets we observe show a mind-blowing diversity that current theories strive to explain as part of the quest to assess the chances of finding life outside the Earth.**

This year's Nobel prize in Physics went to Michel Mayor and Didier Queloz, who discovered the first extrasolar planet orbiting a Sun-like star. 51 Pegasi b, is a gas-giant with a mass approximately half of that of Jupiter. Unlike Jupiter in our Solar System, however, this planet orbits very close to its host star, 100 times closer than Jupiter to our Sun, earning it the classification of "hot Jupiter". One year on 51 Pegasi b lasts only just over 4 days compared to nearly 12 years on Jupiter.

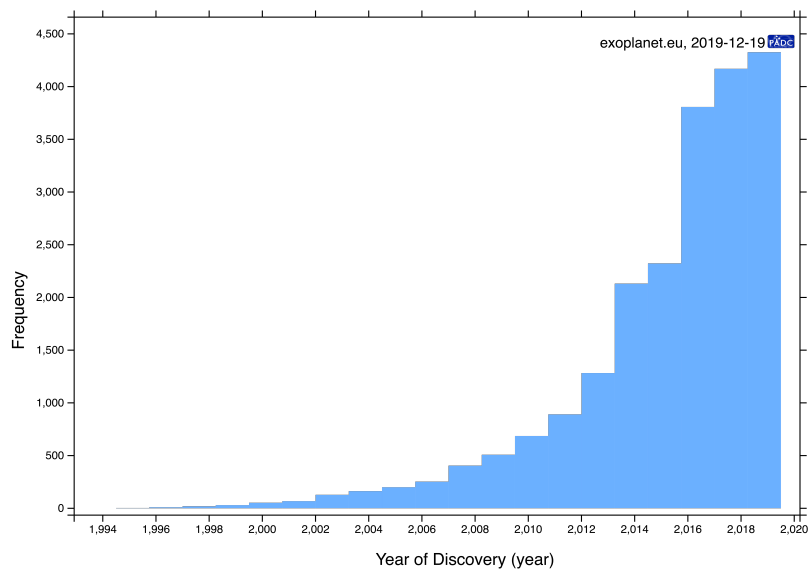

Figure 1: Cumulative distribution of confirmed extrasolar planets. Since the discovery of the first gas giant orbiting a Sun-like star, the rate of discovery has been increasing, with peaks corresponding to the data releases of large space missions likes Kepler.

The combination of large masses and small orbits of hot Jupiters makes them easier to discover through their gravitational interaction with their host stars. The stellar "wobble" induced by the fact that planet and host star orbit around their common centre of mass is detectable via the observation of stellar spectral lines which appear periodically blue- or red-shifted as the planet pulls the star towards and away from us. This observation however requires extremely stable instruments, spectrographs able to resolve the tiny radial velocity component of the star. In the case of 51 Pegasi this velocity is 55 m/s. A planet like Jupiter orbiting the Sun at 5 au is expected to provide a signal of 12.5 m/s, while **the Earth orbiting at 1 astronomical unit (au) from the Sun only pro-**

**duces a signal of 9 cm/s, which is slower than the average walking speed of humans on Earth of roughly 1.5 m/s.** Still detection of Earth-like planets is possible with modern instruments and techniques.

For the detection of the first planet we have to thank the ability of Michel Mayor and his then student Didier Queloz, who were able to perform for the first time the extremely precise measurements required to detect planets and measure their approximate mass. Since 1995, the discovery year of the first planet, this field has been growing exponentially with new planets being discovered at an ever increasing rate via a variety of methods. Today we know of more than 4000 extra-solar planets and it is expected that current missions like the Transiting Exoplanet Survey Satellite (TESS) will discover thousands more via the "transit" method.

During a transit a planets can be detected by the shadow it casts onto the disc of the host star as its orbit crosses our line of sight to the star. By observing the brightness of a star over a long time, the presence of a planet may be thus inferred by the periodic dimming of the stellar light. Space missions like *Kepler* have been extremely successful in detecting planets even as small as the Earth, orbiting their host stars at distances where the temperature on their surface could be suitable for sustaining liquid water at their surface. This so-called circumstellar habitable zone or goldilocks zone is the distance of a planet to its host where the temperature might be just right for life to emerge.

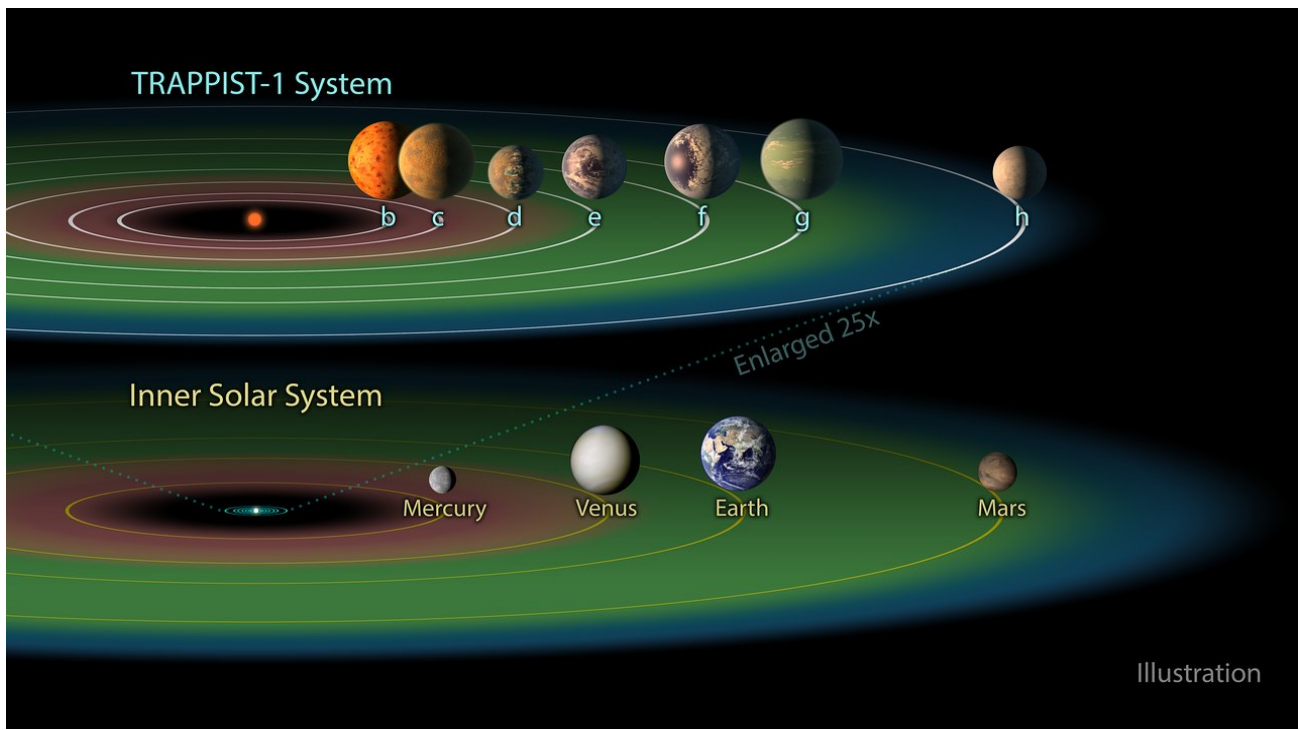

*Figure 2: The Trappist 1 system, discovered by the Kepler mission, which comprises 7 terrestrial planets in the habitable zone of their host star. Credit: NASA/JPL-Caltech*

To date several terrestrial planets with orbits in the habitable zone of their host stars have been discovered. Figure 2 shows an artist impression of the Trappist 1 system, which was initially discovered by the transit method. This system comprises seven terrestrial planets, three of which thought to be in the habitable zone of the host star. However the Trappist 1 system, and many other discovered habitable zone planets orbit stars that are significantly smaller and thus cooler than our Sun, so-called M-stars. Habitable zones of M-stars are very close to the star itself, as the energy required to keep water liquid on the surface of the planet is only available there. This is potentially a problem for life evolving on such planets as tidal locking may occur, which means that the rotation period of the planet around its axis has been synchronised to its revolution period around the star. In this case the planet will always keep the same face towards the star, analogous to the

Earth-Moon system. Temperatures on the opposite sides of tidally locked planets may be very different causing extreme climates. It is currently unclear if life could develop and thrive on tidally locked planets, which do not have a day-night cycle. Perhaps an even more worrying aspect about M-star planet hosts is their high magnetic activity, which produces strong flares of high-energy radiation, which is detrimental to the retention of planetary atmosphere as well as cellular life as we know it on Earth. M-stars which are very abundant in our Galaxy, might not be the best environment to search for signs of life outside our Solar System after all.

A major motivation for exoplanet finding missions is the question of whether the Earth may be a unique and special place for life in our Universe. This question continues to be the driving force behind many observational campaigns and theoretical investigations in the field. While a partial answer may have been provided by the recent exoplanet surveys, which have shown that, **statistically speaking, most stars in the Milky Way have planetary companions** (Cassan et al. 2012), other surveys have also highlighted the diversity of exoplanetary systems (Mullally et al. 2015). This prompted the further question of how did these planets form, and whether some may end up looking like our own and being able to host life.

Diversity and habitability are a consequence of the process of planet formation and subsequent evolution, which is dictated by the physical conditions of the environment in which they form. Planets form from the dust and gas contained in the circumstellar discs surrounding nearly all young low- to intermediate-mass stars. These planet-forming disks are a by-product of the star-formation process, since the gravitational collapse of rotating star-forming clumps necessarily needs to redistribute the angular momentum. So all stars are born with circumstellar discs, meaning that all stars have at least the potential to host a planetary system. Indeed this theoretical prediction is in line with the observation that most stars in our Galaxy host a planet.

Currently, the favourite paradigm for the formation of planets is the so-called core accretion model. In this model micron-size dust grains, embedded in the gaseous circumstellar disc, collide against each other and, in some cases, they stick to each other, growing to larger sizes. This "bottom-up" growth process brings dust grains along a journey over at least 12 orders of magnitude as they reach the sizes of thousands of kilometers. This journey is not easy. Not every collision results in sticking, depending on the velocity of the collision, particles may fragment or just bounce against each other, making the growth process extremely slow. A faster acting, top-to-bottom scenario is disc fragmentation, where the ancestors of planet forming cores emerge as hydrodynamical instabilities in the cooling disc. However planet formation has to happen during the lifetime of the circumstellar disc, which holds the reservoir of material from which planets form.

Circumstellar discs are observed to evolve and finally disperse over a timescale of a few million years, which is comparable to the timescales for planet formation by the core accretion process. This implies that the processes driving the evolution and dispersal of discs play a crucial role in shaping new planetary systems and likely contribute to the observed diversity of exoplanets. Studies show that discs mostly disperse from the inside-out via the formation of so-called transition discs, i.e., discs that have an evacuated inner cavity. These relatively rare objects may be discs caught on the last gasps of their lives and may thus provide key insights on the mechanism responsible for their evolution.

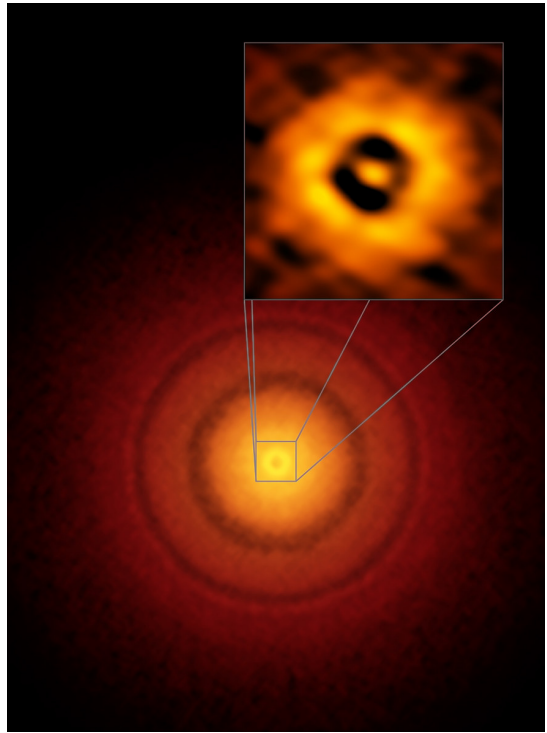

*Figure 3. Atacama Large Millimeter continuum image of TW Hydrae. The outer rings are thought to have been carved by forming planets, while the central cavity has been created by photoevaporation from the central star.*

Disc dispersal is thought to be driven by the high energy radiation, particularly X-ray, emitted from the central star, which plays a key role in shaping the evolution of discs and of their planetary progeny. X-rays penetrate the disc, ionising and heating the gas, which then flows away in a thermally unbound and centrifugally accelerated disc wind. This process of disc photoevaporation ultimately causes the dispersal of the circumstellar disc. The Atacama Large Millimeter Array has provided us with exquisite spatially resolved images of transition discs, where the dispersal process and the planet formation process may be occurring simultaneously as in the case of our nearest neighbour, TW Hydra, shown in Figure 3. The outer disc rings have probably been carved out by forming planets, while the central cavity results from the onset of the photoevaporation process, which will finally cause the demise of the disc and pose an end to the planet formation process in this disc (Ercolano et al. 2017).

The award of this year's Nobel Prize in Physics to exoplanet science is really just the beginning. The era of exoplanet discovery is giving place to a new era of exoplanet characterisation, with new missions, in the near future, designed to determine the physical properties of these far-away worlds as well as starting to study their atmospheric and internal composition, and possible biomarkers (e.g. CHEOPS and Ariel). Biomarkers are gas of biological origins that may hint at the presence of life. At the same time theory is advancing fast to explain how these planets may have formed in their natal circumstellar discs and what role the central star plays in determining the habitability of a system. The next two decades will certainly be full of exciting discoveries in this field, which addresses fundamental questions concerning where we came from and what are the perspectives for the emergence of life elsewhere in the Universe.

About the Author:

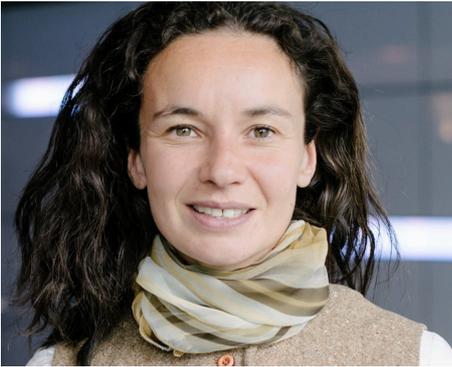

Barbara Ercolano is a Professor of Theoretical Astrophysics at the Ludwig Maximilians University of Munich and leads a DFG Research Unit on transition discs, that are believed to be the natal environments of forming planets. Barbara Ercolano is also a Principal Investigator in the new Origin Excellence Cluster, which is an interdisciplinary Institute joining Astrophysics, Biophysics and Particle Physics in the quest to understand the Origin of the Universe and the building blocks of Life.

She holds a doctoral degree from University College London. She successively held a research position at the Harvard-Smithsonian Center for Astrophysics in Cambridge (MA, USA), an STFC Advanced Fellowship at the Institute of Astronomy of Cambridge University and University College London (UK), and was a lecturer at the University of Exeter (UK). Barbara Ercolano is the 2010 winner of the Fowler Award in Astronomy of the [Royal Astronomical Society](#).